\documentclass[final, 5p]{elsarticle}
\usepackage{amsmath,amssymb}
\usepackage[colorlinks=true,linkcolor=Blue,allcolors=blue]{hyperref}
\usepackage{graphicx}
\usepackage{dcolumn}
\usepackage{bm}
\usepackage{gensymb}
\usepackage{graphicx}
\usepackage{float}
\usepackage{caption}
\usepackage{lipsum}

\newcolumntype{b}{X}
\newcolumntype{s}{>{\hsize=.5\hsize}X}

\journal{Acta Materialia}
\biboptions{sort&compress}

\begin{document}

\begin{frontmatter}

\title{High-throughput screening of Half-antiperovskites with a stacked kagome lattice}

\author[TUD]{Harish K. Singh\corref{mycorrespondingauthor1}}
\cortext[mycorrespondingauthor1]{Corresponding author1}
\ead{harish@tmm.tu-darmstadt.de}


\author[TUD]{Amit Sehrawat}
\author[TUD]{Chen Shen}
\author[TUD]{Ilias Samathrakis}
\author[TUD]{Ingo Opahle}
\author[TUD]{Hongbin Zhang}

\author[TUD]{Ruiwen Xie\corref{mycorrespondingauthor2}}
\cortext[mycorrespondingauthor2]{Corresponding author2}
\ead{ruiwen.xie@tmm.tu-darmstadt.de}

\address[TUD]{Institute of Materials Science, Technical University Darmstadt, Otto-Berndt-Strasse 3, 64287 Darmstadt, Germany}

\date{\today}

\begin{abstract}
Half-antiperovskites (HAPs) are a class of materials consisting of stacked kagome lattices and thus host exotic magnetic and electronic states. We perform high-throughput calculations based on density functional theory (DFT) and atomistic spin dynamics (ASD) simulations to predict stable magnetic HAPs M$_3$X$_2$Z$_2$ (M = Cr, Mn, Fe, Co,
and Ni; X is one of the elements from Li to Bi except noble gases and 4$f$ rare-earth metals; Z = S, Se, and Te), with both thermodynamical and mechanical stabilities evaluated. Additionally, the magnetic ground states are obtained by utilizing DFT calculations combined with the ASD simulations. The existing spin frustration in an AFM kagome lattice manifests as competing behavior of the in-plane FM and AFM couplings. For a total number of 930 HAP compositions considered, we have found 23 compounds that are stabilized at non-collinear antiferromagnetic (AFM) state and 11 compounds that possess ferromagnetic (FM) order. 
\end{abstract}

\begin{keyword}
\texttt Half-antiperovskites \sep Kagome lattice \sep High-throughput \sep Mechanical properties \sep Magnetic ground state 
\end{keyword}

\end{frontmatter}

\section{INTRODUCTION}
\vspace{-0.2cm}
Co$_3$Sn$_2$S$_2$ is a shandite-type structure with a general composition M$_3$X$_2$Z$_2$. Its crystal structure can be derived from antiperovskites (AP) M$_3$XZ with half of the M sites left unoccupied (half antiperovskites, HAP).~\cite{weihrich20113d} A deeper understanding of the structure and bonding in shandite-type compounds can be dated back to around two decades ago.~\cite{weihrich2005halbantiperowskite} Based on the single crystal data, the structure of HAP was refined within space group R$\bar{3}$m (space group number = 166).
With a quasi-two-dimensional crystal structure consisting of stacked kagome lattices (Figure~\ref{fig:Structure}), Co$_3$Sn$_2$S$_2$ is identified as a magnetic Weyl semimetal which hosts a large anomalous Hall conductivity (AHC) and anomalous Hall angle.~\cite{liu2018giant} This discovery triggered huge interest in exploring further the underlying physics and the related intriguing phenomena.~\cite{morali2019fermi,yin2019negative,xu2018topological} 
First it was reported that the ground magnetic state of Co$_3$Sn$_2$S$_2$ is ferromagnetic (FM) with a Curie temperature ($T_c$) of 176 K.~\cite{weihrich2006half} Recently, the existence of an anomalous phase showing magnetic-relaxation-like properties have been observed in the temperature range of $130 < T < 176$ K.~\cite{kassem2017low} Later, a coexistence of FM and antiferromagnetic (AFM) order was evidenced by muon spin-rotation technique.~\cite{guguchia2020tunable} 
The presence of AFM order in Co$_3$Sn$_2$S$_2$ may lead to the magnetic frustration,~\cite{messio2011lattice,boldrin2015haydeeite,boyko2020spin,zhang2021unusual} which is made possible as antiferromagnetically coupled spins are located on the vertices of corner-sharing triangles in the kagome lattice.~\cite{ramirez1994strongly} 

The peculiar geometry of the kagome lattice not only can display strong spin frustration, but also is one of the rare systems showing flat bands.~\cite{yin2019negative,ohgushi2000spin} The flat bands can give rise to a non-zero Chern number originated from a non-trivial Berry phase in the presence of spin-orbit coupling (SOC) and broken time-reversal symmetry, thus making the emergence of interacting topological phases realistic.~\cite{leykam2018artificial} Therefore, the kagome lattice is now acting as a fundamental model in condensed matter physics that hosts exotic magnetic and electronic states.~\cite{balents2010spin,ye2018massive,singh2021multifunctional,singh2022giant}
The key to understanding these exotic magnetic and electronic states is to determine the magnetic ordering of the kagome lattice. Initially, based on the nearest-neighbor Heisenberg model, the low-energy behavior of the kagome antiferromagnet has been extensively studied to seek for spin-liquid phase.~\cite{yan2011spin,sindzingre2009low,hastings2000dirac} It was then realized that further-neighbor interactions also play a crucial role in obtaining physically relevant ground state of materials with kagome geometry.~\cite{janson2008modified,he2015distinct} For instance, the exact diagonalization study for the derived model that includes both the nearest-neighbor (NN) couplings and the couplings across the kagome hexagons has revealed the strong impact of across-hexagon couplings on the magnetic ground state (MGS).~\cite{janson2008modified} 

Despite the significance of gaining more knowledge on the MGSs of HAPs, the evaluation of MGSs is quite challenging due to the difficulty in accurately calculating the exchange interactions. 
Taking Co$_3$Sn$_2$S$_2$ as an example, the out-of-plane NN exchange coupling was suggested to be mainly responsible for the FM order when $T < 130$ K.~\cite{legendre2020magnetic} In stark contrast, Liu et al.~\cite{liu2021spin} concluded the in-plane NN exchange coupling to be the dominant factor. Zhang et al.~\cite{zhang2021unusual} then combined the technique of inelastic neutron scattering with linear spin-wave (SW) theory and found the third-neighbor across-hexagonal exchange as the ruling component for the stabilization of the FM order in Co$_3$Sn$_2$S$_2$. Based on the fitted exchange couplings to SW spectra, Ref.~\cite{zhang2021unusual} speculated that the negligible NN exchange gives rise to the competing AFM and FM interactions in Co$_3$Sn$_2$S$_2$, which are caused by two possible exchange paths, namely Co-S-Co with a bond angle of 76.14° and Co-Sn-Co with a bond angle of 60.04° above and below the kagome layer, respectively. However, density functional theory (DFT) results tend to give rather strong FM exchange coupling for the NN pair.~\cite{zhang2021unusual} Furthermore, as pointed out by Solovyev,~\cite{solovyev2021exchange} the exact theory of exchange interactions, which is beyond the commonly used magnetic force theorem, is required in order to tackle the exact energy change considering, e.g., the contributions from the ligand spins that are expected in HAP. Despite the possible inaccuracies in quantitatively obtaining the exchange coupling parameters, the magnetic force theorem can still provide quite reasonable predictions on the magnetic ground states.~\cite{solovyev2021exchange,zhang2021unusual} Especially for a computationally high-throughput (HTP) screening of the MGSs for HAPs, the use of magnetic force theorem is comparatively more realistic.

In this work, we performed HTP screening for 930 HAP compounds with chemical composition M$_3$X$_2$Z$_2$, where the M atoms (M = Cr, Mn, Fe, Co, and Ni) were selected as magnetic transition metal ions with the aim of finding new magnetic HAPs, X are elements from Li-Bi, except noble gases and 4\textit{f}-elements, and Z atoms are chalcogens (Z = S, Se, and Te). A systematic stability analysis was conducted based on two stability criteria, thermodynamical and mechanical stabilities, determined by the formation energy combined with the convex hull distance and the elastic constants. After validating our calculations using six experimentally known HAPs, we predicted 45 stable novel HAPs. In addition, we evaluated the mechanical properties such as bulk modulus ($\it B$), shear modulus ($\it G$), Young's modulus ($\it E$), and Poisson's ratio ($\it v$) for the stable compounds assuming a FM state. Moreover, the MGS analysis was carried out by comparing the total energies of the FM and non-collinear AFM states (see Figure~\ref{fig:Structure}). Among a total number of 51 stable compounds (45 predicted + 6 known), 35 compounds exhibit magnetism while the rest energetically prefer NM state. On top of that, we identified their MGSs using atomistic spin dynamics (ASD) simulations and evaluated the critical magnetic order transition temperature based on the mean-field (MF) theory.

\vspace{-0.5cm}
\section{Computational Details}
\vspace{-0.2cm}

\subsection{Crystal structure of HAP}

\begin{figure}[htp]
	\begin{center}
	    \includegraphics[width=\columnwidth]{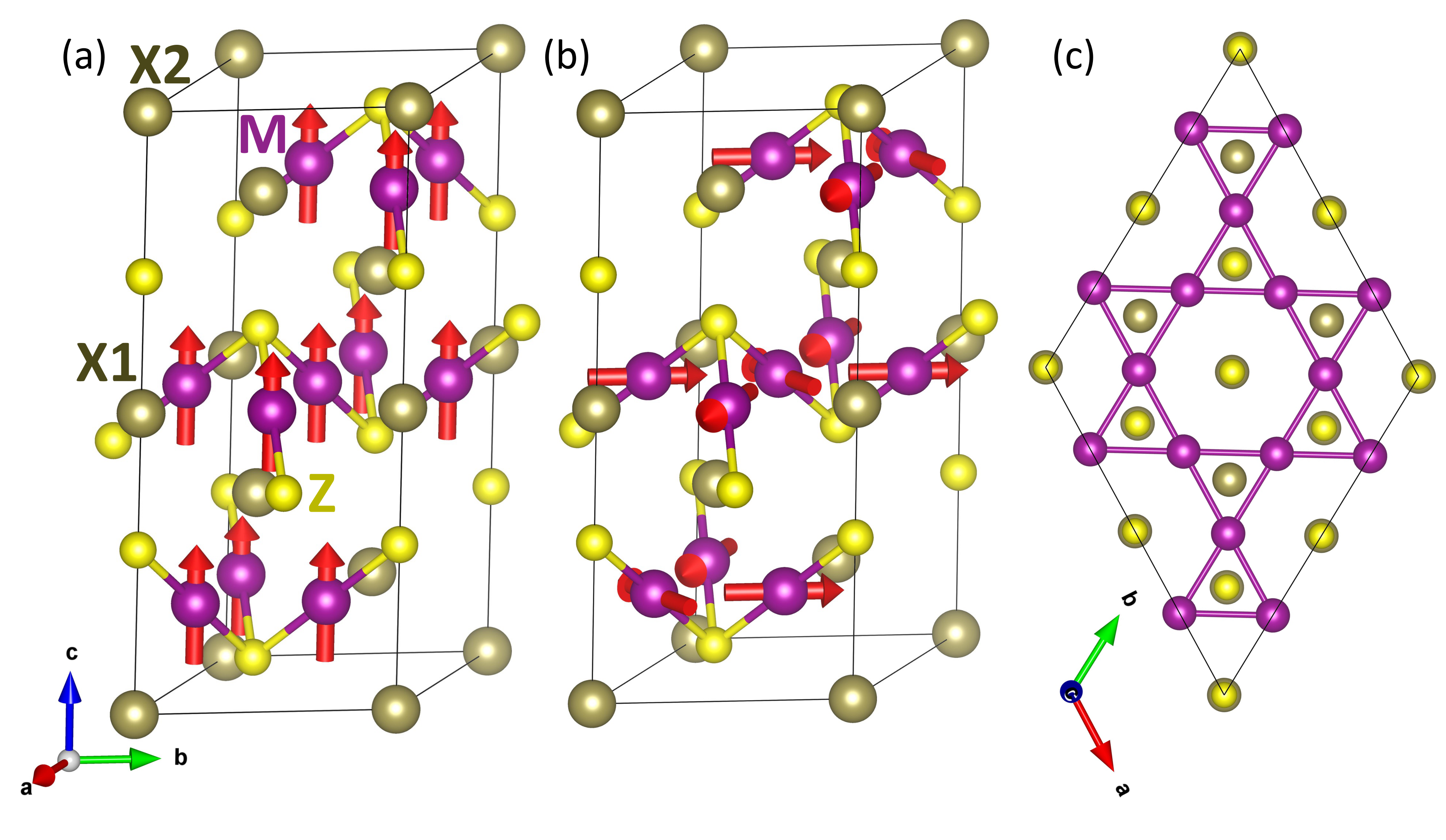}
		\caption{The crystal structure of half-antiperovskites with chemical formula M$_3$X$_2$Z$_2$ in the R$\bar3$m (166) space group, where M atoms form a kagome-lattice. Four Wyckoff positions, including M (0.0,0.5,0.5), X1 (0.0,0.0,0.5), X2 (0.0,0.0,0.0) and Z (0.0,0.0,$z$) are marked out in the figure.} 
		\label{fig:Structure}
	\end{center}
\end{figure} 

HAPs are ternary chalcogenides with a prototypical shandite structure (M$_3$X$_2$Z$_2$), corresponding to the rhombohedral crystal structure within the hexagonal setting, and crystallize in the R$\bar{3}$m space group (space group number = 166) (Figure~\ref{fig:Structure}). In a trigonal rhombohedral structure, the M atoms occupy the 9e (1/2,0,0) sites, while X atoms hold two independent positions on 3a (0,0,1/2) and 3b (0,0,0), whereas the Z atoms are located at the 6c position (0,0,c). Note that except the Z atoms, the atomic positions of M and X are fixed. A characteristic feature of the shandite structure is that the M atoms form a 2D-sheet of kagome sublattice in each stacking layer along c-axis, constituting a topology of ABC pattern (see Figure~\ref{fig:Structure}). The M atoms are octahedrally coordinated with two Z atoms and four X2 atoms, where the X2 and Z atoms are located between the layers. Each M$_3$ kagome sublattice is capped above and below by a X2 and a Z atom producing a trigonal anti-prismatic inter-layer sites. The X1 atoms are amalgamated within the 2D kagome-layer, coordinated with the M atoms forming hexagonal-planar structure. 

\subsection{Stability evaluation}

\begin{figure}[ht!]
	\begin{center}
	    \includegraphics[width=0.85\columnwidth]{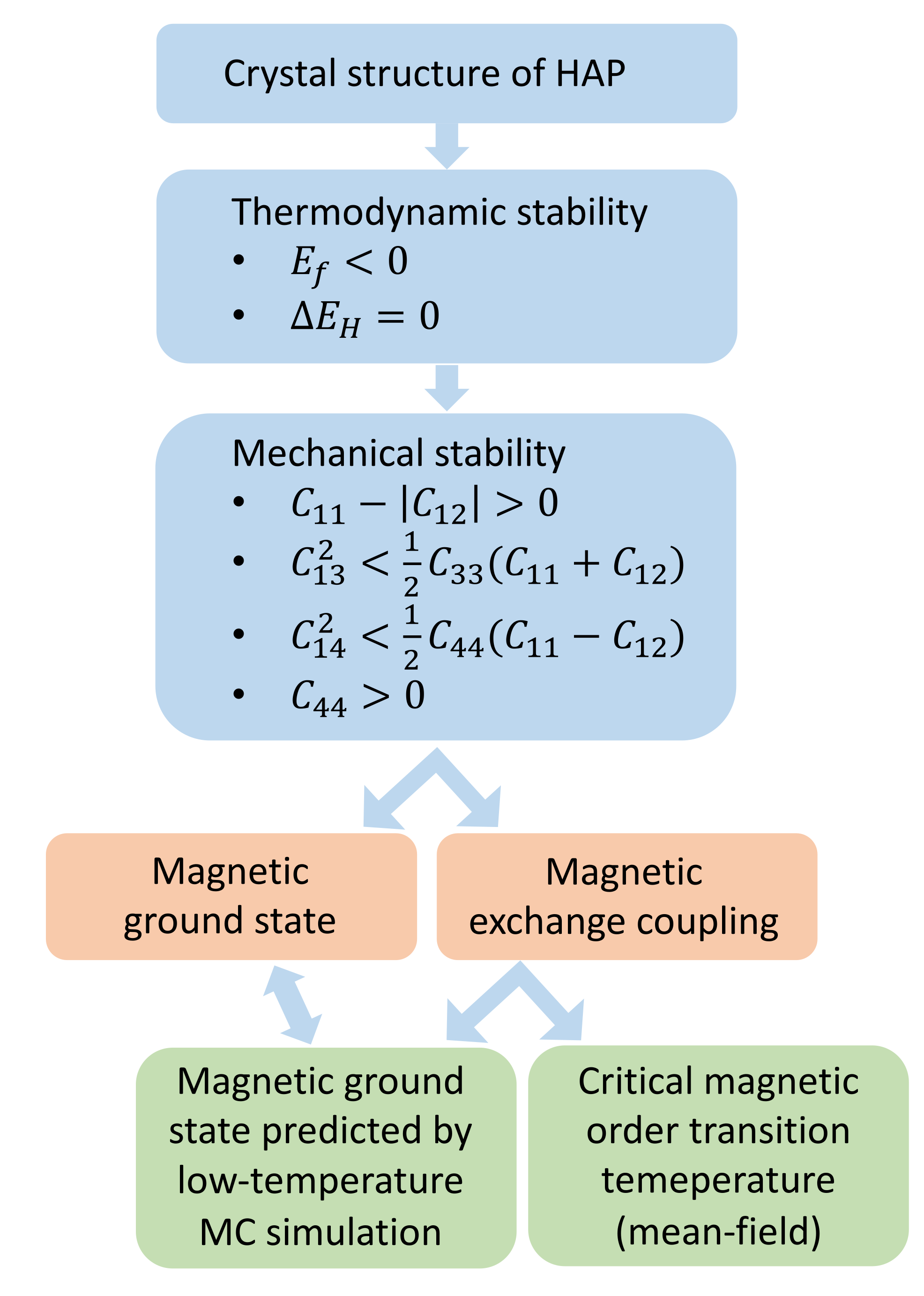}
		\caption{Workflow for the high-throughput screening.}
		\label{fig:workflow}
	\end{center}
\end{figure}

The high-throughput screening process is summarized in the workflow as shown in Figure~\ref{fig:workflow}. The DFT calculations were conducted using an in-house developed high-throughput environment to evaluate the thermodynamic stability ~\cite{opahle2012high,opahle2013high,Singh2018,opahle2020effect,shen2021designing}, which is determined in two steps. The first step is the formation energy ($E_f$), which can be defined as $E_f$ = $E_{M_3X_2Z_2}$ - (3$E_M$ + 2$E_X$ + 2$E_Z$), where $E_{M_3X_2Z_2}$ is the total energy of the HAP compound, and $E_M$, $E_X$, and $E_Z$ are the total energies of the constituent elements in their bulk phase. In the second step, the convex hull distance ($E_H$) is evaluated by considering all the competing phases that exist in the open quantum materials database (OQMD).~\cite{kirklin2015open} $E_H$ for a given phase can be defined as: $\Delta E_H$ = $E_{tot}$ (given phase) - $E_{tot}$ (competing phase). For the compounds, $E_H$ $<$ 100 meV/atom, the mechanical stability and mechanical properties were calculated using the code developed by Singh et al.~\cite{singh2018elastic} The mechanical properties were determined by utilizing the Voigt~\cite{voigt1966lehrbuch} and Reuss~\cite{reuss1929calculation} schemes which provide an upper and lower bound estimation of the elastic moduli, respectively. Thereby, an arithmetic average was taken into consideration according to the Voigt-Reuss-Hill approximation (see Table~\ref{tab:table1}).~\cite{hill1952elastic} 
Note that the thermodynamical and mechanical stabilities were examined assuming the FM state, which is sufficient for stability analysis. 

The DFT calculations were conducted using the projector augmented wave (PAW) method as implemented in the Vienna \textit{ab} \textit{initio} simulation package (VASP).~\cite{kresse1996} A generalized gradient approximation (GGA) was used to approximate exchange-correlation functional under the parameterized form of Perdew-Burke-Ernzerhof (PBE).~\cite{perdew1996generalized} The energy cutoff was set at 520 eV. A consistent k-mesh of 13$\times$13$\times$5 was used within the Monkhorst-pack scheme, and a Methfessel-Paxton smearing width was set to 0.06 eV. The crystal structures were optimized until the maximum residual force at each atom was less than 0.001 eV\AA$^{-1}$. The optimized lattice parameters are in good agreement with the experimentally reported parameters (Table~\ref{tab:table1}). The thermodynamical stability trends of HAPs were elucidated by analyzing the bond strength using the Crystal Orbital Hamilton Population (COHP) approach as implemented in the Lobster code.~\cite{Maintz2016}

\subsection{Magnetic ground state and critical magnetic order transition temperature evaluation \label{sec:method:MC}}
For HAP systems that are thermodynamically and mechanically stable, as well as magnetic, we examined further their MGSs by taking into account one FM and one non-collinear AFM configuration (see Fig~\ref{fig:Structure} (a) and (b)). Moreover, considering the rich nature of the magnetic configurations in the kagome lattice, we also validated the MGSs using ASD simulations. 
Here, we would like to mention that the MGS predicted by ASD simulation is, on the one hand, a justification of the DFT result if both ASD and DFT show FM as ground state. On the other hand, it can be expected that the ASD simulations might give AFM state as ground state in contrast to the FM ground state evaluated by DFT. The reason for this is that it is almost impossible for DFT calculations to cover all possible AFM configurations. 
Another point is that for M$_3$X$_2$Z$_2$ predicted to be AFM using both DFT and ASD simulations, the AFM configuration given by ASD could be different from the one shown in Fig~\ref{fig:Structure} (b). 
At last, we would like to stress that it is a rather formidable task to find the true MGS of an AFM kagome lattice which is non-Bravais.~\cite{messio2011lattice} In the present work, for the evaluation of the MGS, we have to compromise between computational cost and accuracy in a high-throughput procedure. Our aim is thus to provide an initial prediction on the magnetic nature of the kagome lattice, being FM or AFM. Detailed investigation on the specific magnetic configurations corresponding to the AFM state is left for future work.

The critical magnetic temperature was evaluated with the mean-field approximation,~\cite{csacsiouglu2004first,anderson1963theory} where the interatomic exchange parameters, defined in terms of the classical Heisenberg Hamiltonian, were calculated using the OpenMX code.~\cite{ozaki2003variationally,ozaki2004numerical,ozaki2005efficient} The selected local basis sets for various elements are listed in Supplementary Information (SI) Table S1. The exchange-correlation functional of GGA-PBE was adopted. The exchange coupling parameters were calculated based on the Lichtenstein formula~\cite{liechtenstein1987local} with respect to the FM state using the post-processing code ‘jx’ in OpenMX~\cite{han2004electronic}. The energy cut-off and the k-mesh were set to be 400 Ry and 40$\times$40$\times$20, respectively. The self-consistent energy convergence criterion was chosen to be 1.0$\times$e$ ^{-8} $ Hartree. 
The MGS was examined at 10 K by inputting the calculated interatomic exchange parameters into Uppsala Atomistic Spin Dynamics (UppASD) program~\cite{eriksson2017atomistic} implementing the Metropolis Monte Carlo (MC) algorithm~\cite{landau2021guide}.

\section{RESULTS and DISCUSSION \label{sec:res}}
\begin{table*}
    \footnotesize
    \centering
	\caption{\label{tab:table1}For HAP compounds M$_3$X$_2$Z$_2$ on the covex hull ($\Delta E_H=0$), the various physical properties are summarized, such as the formation energy ($E_f$), bulk modulus ($\it B$), shear modulus ($\it G$), Young's modulus ($\it E$), Poisson's ratio ($\it v$), the magnetic ground state (MGS) given by DFT and ASD simulations, and the critical magnetic order transition temperature ($T_C$ or $T_N$). The experimentally reported compounds are marked with an asterisk.}
	\begin{tabular}{lccccccccccc}
	            \hline
				&M$_3$X$_2$Z$_2$&\multicolumn{2}{c}{Lattice constant}&E$_f$&$\it B$&$\it G$&$\it E$&$\it v$ & MGS & MGS & $T_C/T_N$\\
				S.No.&Compounds&a (\footnotesize{\AA})&c (\footnotesize{\AA})&eV/atom&GPa&GPa&GPa& &DFT&ASD& MF (K)\\ 
				\hline\centering
				(1)&Co$_3$In$_2$S$_2$* &5.326 &13.643 &  -0.6008 &82.7&50.2&125.3&0.25 & AFM & AFM & 54 \\
				&&(5.317)&(13.666)&&&&&&& PM~\cite{mcguire2021antiferromagnetic} & \\
				(2)&Co$_3$Sn$_2$S$_2$*&5.385 &13.176 &  -0.6238  &93.8 (97)&53.0  (57)&133.9&0.26 (0.26) & FM & FM & 82 \\
				&&(5.357)&(13.127)&&&&&&&&(167~\cite{liu2021spin}) \\
				(3)&Co$_3$Ga$_2$S$_2$  &5.140&12.836&  -0.6257 &91.6&59.5&146.8&0.23 & AFM & AFM & 22\\
				(4)&Co$_3$Ge$_2$S$_2$  &5.195&12.460&  -0.5386 &102.1&61.8&154.4&0.25 & FM & FM & 43 \\
				(5)&Co$_3$Pb$_2$S$_2$  &5.495&13.719&  -0.4645 &74.5&39.2&100.2&0.28 & FM & FM & 329\\
				(6)&Co$_3$Bi$_2$S$_2$  &5.655&13.225&  -0.4481 &76.7&43.0&108.8&0.27 & FM & AFM & 219\\
				(7)&Co$_3$Sb$_2$S$_2$  &5.531&12.770&  -0.5100 &75.6&52.8&128.5&0.22 & FM & AFM & 294\\
				(8)&Co$_3$Tl$_2$S$_2$  &5.413&14.052&  -0.4192 &54.5&40.7&97.9&0.20 & FM & FM & 266 \\
				(9)&Co$_3$Al$_2$S$_2$  &5.101&12.567&  -0.7379 &63.3&67.7&149.8&0.11 & NM & & \\
				(10)&Fe$_3$In$_2$S$_2$ &5.359&14.132&  -0.4191 &63.1&43.2&105.5&0.22 & FM & AFM & 1098 \\
				(11)&Fe$_3$Sn$_2$S$_2$ &5.341&13.393&  -0.4260 &103.7&59.4&149.6&0.26 & FM & FM & 343\\
				(12)&Mn$_3$In$_2$S$_2$ &5.455&14.268&  -0.5272 &77.4&47.0&117.3&0.24&FM & AFM & 231\\
				(13)&Mn$_3$Sb$_2$S$_2$ &5.665&13.336&  -0.4996 &49.1&43.2&100.2&0.16&FM & AFM & 533\\
				(14)&Mn$_3$Ga$_2$S$_2$ &5.266&13.450&  -0.5613 &76.8&50.1&123.3&0.233 & FM & FM & 374 \\
				(15)&Mn$_3$Sn$_2$S$_2$ &5.417&13.706&  -0.5677 &89.4&62.1&151.3&0.22& FM & AFM & 495\\
				(16)&Mn$_3$Te$_2$S$_2$ &5.960&13.934&  -0.5057 &24.6&15.4&38.4&0.24 & AFM & AFM & 308\\
				(17)&Mn$_3$Bi$_2$S$_2$ &5.839&13.8334&  -0.4276 &44.1&35.8&84.6&0.18 & FM & FM & 624 \\
				(18)&Mn$_3$Pb$_2$S$_2$ &5.829&14.244&  -0.4308 &37.1 &25.1 &61.5&0.22 & AFM & AFM & 460\\
				(19)&Mn$_3$Se$_2$S$_2$ &6.114&12.238&  -0.5845 &30.5&19.2&47.6&0.24 & AFM & AFM & 277\\
				(20)&Mn$_3$Tl$_2$S$_2$ &5.547&14.573&  -0.3997 &42.9&32.0&76.9&0.20 & AFM & AFM & 718\\
				(21)&Mn$_3$Hg$_2$S$_2$ &5.612&14.734&  -0.3817 &40.5&22.0&55.9&0.27 & AFM & AFM & 832\\
				(23)&Mn$_3$As$_2$S$_2$ &5.486&12.527&  -0.4788 &64.6&50.3&119.9&0.19 & AFM & AFM & 313\\
				(24)&Ni$_3$In$_2$S$_2$* &5.435 &13.639 &  -0.5893 &62.0 &31 &78.6&0.29  & NM & & \\
				&&(5.369)&(13.571)&&(72)& (33) &&(30)&&&\\
				(25)&Ni$_3$Pb$_2$S$_2$* &5.687 &13.789 &  -0.5021 &43.2&22.6&57.7&0.28 & NM & & \\
				&&(5.576)&(13.658)&&&&&&&&\\
				(26)&Ni$_3$Cd$_2$S$_2$&5.367&14.065&  -0.5192 &&&& &NM & & \\
				(27)&Ni$_3$Sn$_2$S$_2$* &5.551 &13.231&  -0.5584 &53.2&33.1&82.3&0.24 & NM & &\\
				&&(5.471)& (13.209)&&&&&&&&\\
				(28)&Ni$_3$Hg$_2$S$_2$ &5.426&14.177&  -0.4064 &48.9&23.7&61.3&0.29& NM & & \\
				(29)&Ni$_3$Bi$_2$S$_2$ &5.818&13.186&  -0.4274 &52.3&25.7&66.4&0.29 & NM & &\\
				(30)&Ni$_3$Ge$_2$S$_2$ &5.355&12.607&  -0.4546 &49.2&33.0&81.0&0.23 & NM & &\\
				(31)&Ni$_3$Sb$_2$S$_2$ &5.698&12.763&  -0.4392 &36.8&30.4&71.6&0.17 & NM & & \\
				(32)&Ni$_3$Ga$_2$S$_2$ &5.221&12.981&  -0.5719 &51.5&32.2&79.9&0.24 & NM & &\\
				(33)&Ni$_3$Ag$_2$S$_2$&5.291&14.044& -0.7467   &49.2&26.0&66.4&0.27 & NM & &\\
				(34)&Co$_3$Sn$_2$Se$_2$&5.445&13.805& -0.4031  &86.9&50.2&126.3&0.26 & FM & FM & 80\\
				(35)&Co$_3$Bi$_2$Se$_2$&5.701&13.872& -0.2572  &62.1&36.8&92.2&0.25 & FM & AFM & 167\\
				(36)&Co$_3$Sb$_2$Se$_2$&5.578&13.451& -0.3044  &25.3&41.6&80.4&-0.04 & FM & FM & 263\\
				(37)&Co$_3$Ga$_2$Se$_2$&5.202&13.567& -0.3868  &24.8&48.9&88.6&-0.10 & FM & AFM & 48 \\
				(38)&Co$_3$Ge$_2$Se$_2$&5.233&13.285& -0.2935  &90.3&54.6&136.4&0.25 & NM & &\\
				(39)&Mn$_3$Sb$_2$Se$_2$&5.746&14.066& -0.3141  &45.0&35.7&84.7&0.18& FM & AFM & 464\\
				(40)&Mn$_3$Sn$_2$Se$_2$&5.502&14.487& -0.3604  &59.8&46.4&110.6&0.19 & FM & AFM & 290 \\
				(41)&Mn$_3$Te$_2$Se$_2$&5.943&14.745& -0.3380  &25.6&15.0&37.6&0.25 & AFM & AFM & 323\\
				(42)&Mn$_3$Bi$_2$Se$_2$&5.904&14.486& -0.2656  &42.8&31.7&76.4&0.20 & FM & FM & 509\\
				(43)&Mn$_3$Hg$_2$Se$_2$&5.736&15.288& -0.2153  &18.5&17.3&39.7&0.14 & AFM & AFM & 855 \\
				(44)&Mn$_3$Tl$_2$Se$_2$&5.902&15.661& -0.2407  &26.0&17.7&43.4&0.22 & AFM & AFM & 353\\
				(45)&Mn$_3$In$_2$Se$_2$&5.522&14.849& -0.3402  &57.1&36.6&90.4&0.24 & AFM & AFM & 464\\
				(46)&Mn$_3$Ga$_2$Se$_2$&5.322&14.176& -0.3486  &79.0&46.2&116.0&0.26 & FM & FM & 217\\
				(47)&Mn$_3$Se$_4$&6.385&12.765& -0.4023  &35.5&17.0&44.0&0.29 & AFM & AFM & 311 \\
				(48)&Ni$_3$Hg$_2$Se$_2$&5.453&14.876& -0.2465  &8.6&22.7&36.2&-0.21 & NM & & \\
				(49)&Ni$_3$Sn$_2$Se$_2$&5.572&13.992& -0.3929  &37.3&28.3&67.9&0.20 & NM & &\\
				(50)&Ni$_3$Bi$_2$Se$_2$&5.810&13.937& -0.2797  &38.6&26.4&64.5&0.22 & NM & & \\
				(51)&Ni$_3$Pb$_2$Se$_2$*&5.717 &14.494 & -0.3516  &48.9&23.9 &61.7&0.30  & NM & &\\
				&&(5.617)&(14.286)&& (55)&(22)&&(0.32)&&&\\
				(52)&Ni$_3$Tl$_2$Se$_2$&5.577&14.869& -0.2973  &22.0&18.8&44.0&0.17 & NM & & \\
				\hline
			\end{tabular}
\end{table*}

\subsection{Thermodynamical stability}
Evaluating material thermodynamical stability, which is determined by the formation energy ($E_f$) and the convex hull distance ($\Delta E_H$), is critical in high-throughput materials design and is one of the key criteria for predicting new materials. A compound is thermodynamically stable as long as $E_f \textless 0$ and $\Delta E_H=0$ conditions are fulfilled. In general, $E_f$ is an indispensable but insufficient condition, because a negative $E_f$ only ensures stability against decomposition into its constituent elements, whereas $\Delta E_H=0$ guarantees stability against decomposition into any possible combination of the competing phases. A given phase with $\Delta E_H=0$ is the most stable phase against decomposition and part of the convex hull, while a given phase with $\Delta E_H$ $>$ 0 is considered plausibly stable (metastable) up to a threshold value. For metastable phases, we have chosen a tolerance of $\Delta E_H$ $\leq$ 100 meV/atom based on consideration of approximations in the exchange correlation functional, numerics, and temperature effects. A threshold range of 30-100 meV/atom of $\Delta E_H$ is considered in various high-throughput studies~\cite{Singh2018,shen2021designing,gao2019high,opahle2013high}.\par
We first validated our high-throughput calculations for the thermodynamical stability criteria for the six experimentally reported HAP compounds that are found in the thorough literature search (Table~\ref{tab:table1}). All experimentally reported HAPs have negative formation energies ($E_f$ $\textless$ 0) and are part of the convex hull with $E_H = 0$, which means that all six compounds are thermodynamically stable. Moreover, the calculated formation energies for the compounds Ni$_3$In$_2$S$_2$, Ni$_3$Pb$_2$S$_2$, and Ni$_3$Sn$_2$S$_2$ agree well with those reported in the Materials Project (MP) database. On the other hand, differences in the formation energies of 0.078, -0.091, and 0.109 eV/atom are observed for Co$_3$Sn$_2$S$_2$, Co$_3$In$_2$S$_2$, and Ni$_3$Pb$_2$Se$_2$, respectively, which is due to the high density k-mesh in our calculations. In addition, the determined distance to the convex hull from our study agrees well with the MP database with $E_H = 0$, except for Ni$_3$Pb$_2$S$_2$, where it is on the convex hull according to our calculations ($E_H=0$) while the MP database value is 4 meV/atom.\par
\begin{figure}[htp]
	\begin{center}
	    \includegraphics[width=\columnwidth]{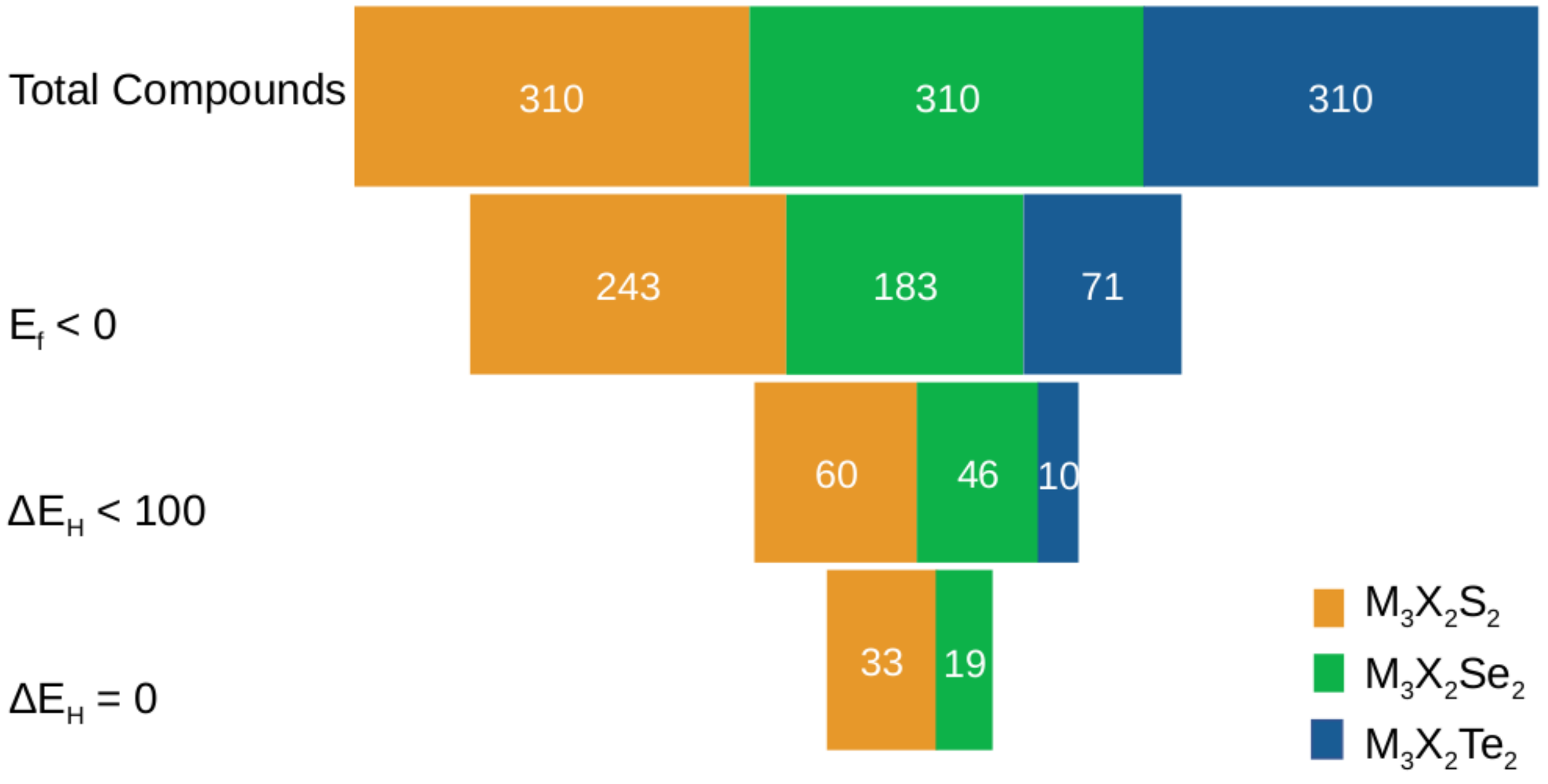}
		\caption{Thermodynamical stability flow diagram for the M$_3$X$_2$Z$_2$ (Z = S, Se, and Te), showing a decrease in stable compounds with more restrictive stability criteria E$_f$, $\Delta E_H$ $<$ 100 meV, and $\Delta E_H$ = 0.}
		\label{fig:Stability}
	\end{center}
\end{figure}

As shown in Figure~\ref{fig:Stability}, the number of compounds with a negative $E_f$ value decreases from 243 to 183 to 71 when Z in the chemical formula M$_3$X$_2$Z$_2$ is substituted from S to Se to Te. A similar trend is observed for the convex hull distance (Figure~\ref{fig:Stability}). A total of 33 and 19 compounds are part of the convex hull ($\Delta E_H = 0$) consisting of S- and Se-based HAPs, respectively, and none of the Te-based HAPs has $\Delta E_H=0$. To understand the stability trend of $E_f$ and $\Delta E_H$, chemical bond analysis was performed. The integrated chemical orbital Hamilton population (ICOHP) was calculated using the LOBSTER code~\cite{Lobster,burghaus2010ternary} to identify the chemical bond responsible for the stability. For this reason, we selected Mn$_3$Ga$_2$S$_2$, Mn$_3$Ga$_2$S$_2$, and Mn$_3$Ga$_2$Te$_2$ compounds demonstrating decreasing $E_f$ values of -0.56, -0.35, and -0.05 eV/atom, respectively. The first two compounds are on the convex hull ($\Delta E_H$ =0), and the last compound is above the convex hull ($\Delta E_H$ = 0.159 eV/atom). The bonds between Mn-Mn, Mn-Z (Z = S, Se and Te), Mn-Ga, Ga-Ga and Z-Z atoms were considered for the bond strength analysis. A smaller value of average ICOHP corresponds to the more stable compounds. The spin-up (spin-down) average ICOHP values for Mn$_3$Ga$_2$S$_2$, Mn$_3$Ga$_2$S$_2$, and Mn$_3$Ga$_2$Te$_2$ are -2.61 (-2.91), -2.59 (-2.81), and -2.13 (-2.32), respectively, which agrees well with the trends of $E_f$ and $\Delta E_H$. On moving down the group in the periodic table, the ionic radii increase from S (1.84 \AA) to Se (1.98 \AA) to (2.21 \AA). As a result, the bond distances between Mn-S, Mn-Se, and Mn-Te atoms also increase by 2.22, 2.35, and 2.73 \AA, respectively. Their bond-resolved ICOHP values are -1.45, 1.43, and -1.15, respectively, which contribute most to the stability of the corresponding compounds. The bonds between Z-Z atoms are the weakest, contributing least to the stability of the compounds. 
\\
\\
\vspace{-0.6cm}
\subsection{Mechanical stability}  
The compounds are likely to be formed if they are thermodynamically stable. However, it does not corroborate a further decrease in total energy against structural deformation or distortion upon the application of mechanical strain, which can be assessed by the mechanical stability analysis. A compound is mechanically stable as long as it has absolute minimum energy for the unstrained state. Therefore, mechanical stability is formulated in terms of strain energy (W), which can be calculated from the energy difference between the strained ($E_{str}$) and the unstrained ($E_{unstr}$) states based on the elastic constants.
\begin{equation}
\label{eq:MS}
W = E_{str} - E_{unstr} = \frac{1}{2} \sum_{i,j} C_{i,j} \epsilon_{i}\epsilon_{j},
\end{equation}  
where $\epsilon$ denotes strain and $C_{i,j}$ are second-order elastic tensors. For crystals with rhombohedral Laue class $\bar{3}$m, there are six independent elastic constants: $C_{11}$, $C_{12}$, $C_{13}$, $C_{14}$, $C_{33}$, and $C_{44}$, where, $C_{66} = (C_{11} - C_{12})/2$ as in the hexagonal case. Therefore, the elastic tensor is expressed as follows:
\begin{equation}
\begin{pmatrix}
C_{11}&C_{12}&C_{13}&C_{14}&0&0\\
C_{12}&C_{11}&C_{13}&-C_{14}&0&0\\
C_{13}&C_{13}&C_{33}&0&0&0\\
C_{14}&-C_{14}&0&C_{44}&0&0\\
0&0&0&0&C_{44}&C_{14}\\
0&0&0&0&C_{14}&C_{66}\\
\end{pmatrix}
\end{equation} 
A compound is mechanically stable if it fulfills the Born-Huang stability criteria.~\cite{born1955dynamical,mouhat2014necessary} According to such criteria, the strain energy (W) should be positive, i.e., all eigenvalues of the C$_{ij}$ matrix must be positive. For rhombohedral Laue class $\bar{3}$m, the mathematical expressions for Born stability have been outlined by Mouhat et al.,~\cite{mouhat2014necessary} which are defined as
\begin{subequations}
\begin{align}
\label{eq:Born1}
C_{11} - |C_{12}| > 0, 
\\
\label{eq:Born2}
C^2_{13} < \frac{1}{2}C_{33}(C_{11}+C_{12}),
\\
\label{eq:Born3}
C^2_{14} < \frac{1}{2}C_{44}(C_{11}-C_{12})\equiv C_{44}C_{66},
\\
\label{eq:Born4}
C_{44} > 0.
\end{align}
\end{subequations}

The mechanical stability is analyzed for experimentally reported HAP and all predicted compounds with a convex hull distance $<$100 meV. All six experimentally synthesized compounds (marked with asterisk Table~\ref{tab:table1}) fulfill the Born stability criteria and are therefore mechanically stable. For example, Co$_3$Sn$_2$S$_2$ fulfills all conditions for mechanical stability. For the predicted compounds with $\Delta E_H$ = 0, all compounds fulfill stability criteria except Ni$_3$Cd$_2$S$_2$, where Born stability conditions~\ref{eq:Born2}, and~\ref{eq:Born3} are not fulfilled, thus the system becomes mechanically unstable. For the systems above the convex hull (0 $<$ $\Delta E_H$ $<$100 meV), eight out of sixty-four compounds are mechanically unstable (see SI Table S2).

\subsection{Mechanical properties} 
Next, we evaluated various mechanical properties including bulk modulus ($\it B$), shear modulus ($\it G$), Young's modulus ($\it E$), and Poisson's ratio ($\nu$), which were obtained using the elastic tensors $C_{i,j}$. The mechanical properties are briefly defined below. The bulk modulus ($\it B$) quantifies the strength of a compound to resist volume compressibility under mechanical stress from all sides and the shear modulus ($\it G$) measures the strength of a compound to withstand transverse deformation. Pugh proposed~\cite{pugh1954xcii} that solids' brittle and ductile behavior can be analyzed by the ratio of bulk modulus to shear modulus (B/G). The key value to distinguish between brittle and ductile materials is 1.75 and a $\it B$/$\it G$ value below/above 1.75 represents the brittle/ductile character of the solids. The Young's modulus ($\it G$) is defined as ratio of stress to strain along the same axis, i.e., it measures the stiffness of the materials, which can be calculated using the following expression: $\it E$ = 9$\it B$$\it G$$/$(3$\it B$+$\it G$). Our calculated elastic constants and the corresponding mechanical properties agree well with the values reported in the MP database (see Table~\ref{tab:table1}).~\cite{de2015charting} For instance, the values of $\it B$, $\it G$, and $\nu$ for Co$_3$Sn$_2$S$_2$ from our calculations (MP database) are 94 (97), 53 (57), and 0.26 (0.26), respectively.\par

\begin{table*}[ht!]
    \centering
	\caption{\label{tab:mag} Comparison of exchange couplings and mean-field $T_C$ of Co$_3$Sn$_2$S$_2$ using different approaches. In the current notation, a positive (negative) sign indicates ferromagnetic (antiferromagnetic) coupling.}
	\begin{tabular}{lcccccccccc}
	            \hline
				& $J_1$ & $J_2$& $J_3$ & $J_3^{\prime}$ &$J_{c1}$ & $J_{c2}$& $J_{c2}^{\prime}$ & $J_{c3}$ &T$_C$ & Reference \\
				Approach & (meV) & (meV) & (meV) & (meV) & (meV) & (meV) & (meV) & (meV) &  (K)& \\ 
				\hline\centering
				OpenMX & 1.08 & -0.01 & 0.16 & 0.08 & 0.06 & 0.55  & 0.22 & 0.23 & 81 & this work\\ 
				OpenMX (GGA + U) & 2.12 & -0.11 & 0.25 & 0.05 & 0.03 & 0.95  & 0.29 & 0.31 & 127 & this work\\ 
				SPR-KKR (GGA + U) & 1.26 & -0.02 & 0.22 & 0.62 & 0.06 & 0.86 & 0.86 & 0.60 & 130 & Ref.~\cite{zhang2021unusual}\\ 
				Fits to SW spectra &  & -0.98 &  & 5.09 & 1.97 & 0.29 &  &  & 175 & Ref.~\cite{zhang2021unusual}\\ 
				MFWF & 2.36 & 0.16 & 0.02 & 0.02 & 0.30 & 0.75 & 0.75 & 0.11 & 167 & Ref.~\cite{liu2021spin} \\ 
			    \hline
	\end{tabular}
\end{table*}

As shown in Table~\ref{tab:table1}, the HAP compounds encompass a broad range of mechanical properties. For the HAP compounds with $\Delta E_h=0$, the bulk modulus ($\it B$) ranges from 8.6 GPa in Ni$_3$Hg$_2$Se$_2$ to 103.7 GPa in Fe$_3$Sn$_2$S$_2$ and the shear modulus ($\it G$) varies from 15 GPa in Mn$_3$Te$_2$Se$_2$ to 67.7 GPa in Co$_3$Al$_2$S$_2$. Moreover, 11 out of 52 are ductile (see Table~\ref{tab:table1}). For instance, Co$_3$Sn$_2$S$_2$ is ductile with a $\it B$/$\it G$ value of 1.77, exhibiting a partial ductility character, while Ni$_3$Pd$_2$Se$_2$ displays an assertive ductile behavior reaching a maximum $\it B$/$\it G$ value of 3.1. Finally, the Young's modulus ($\it E$) fluctuates from 37.6 GPa in Mn$_3$Te$_2$Se$_2$ to 154.4 GPa in Co$_3$Ge$_2$S$_2$. The larger magnitude of the Young modulus implies a larger stiffness tendency of the materials. For the compounds above the convex hull, Co$_3$Hf$_2$S$_2$ ($\Delta E_h=39.4$ meV/atom) attains the maximum value of 147.2 GPa, 83 GPa, and 209.7 GPa for the bulk modulus ($\it B$), shear modulus ($\it G$) and Young's modulus ($\it E$), respectively. 
 
\begin{figure*}[h!]
\centering
	\begin{center}
	    \includegraphics[width=\linewidth]{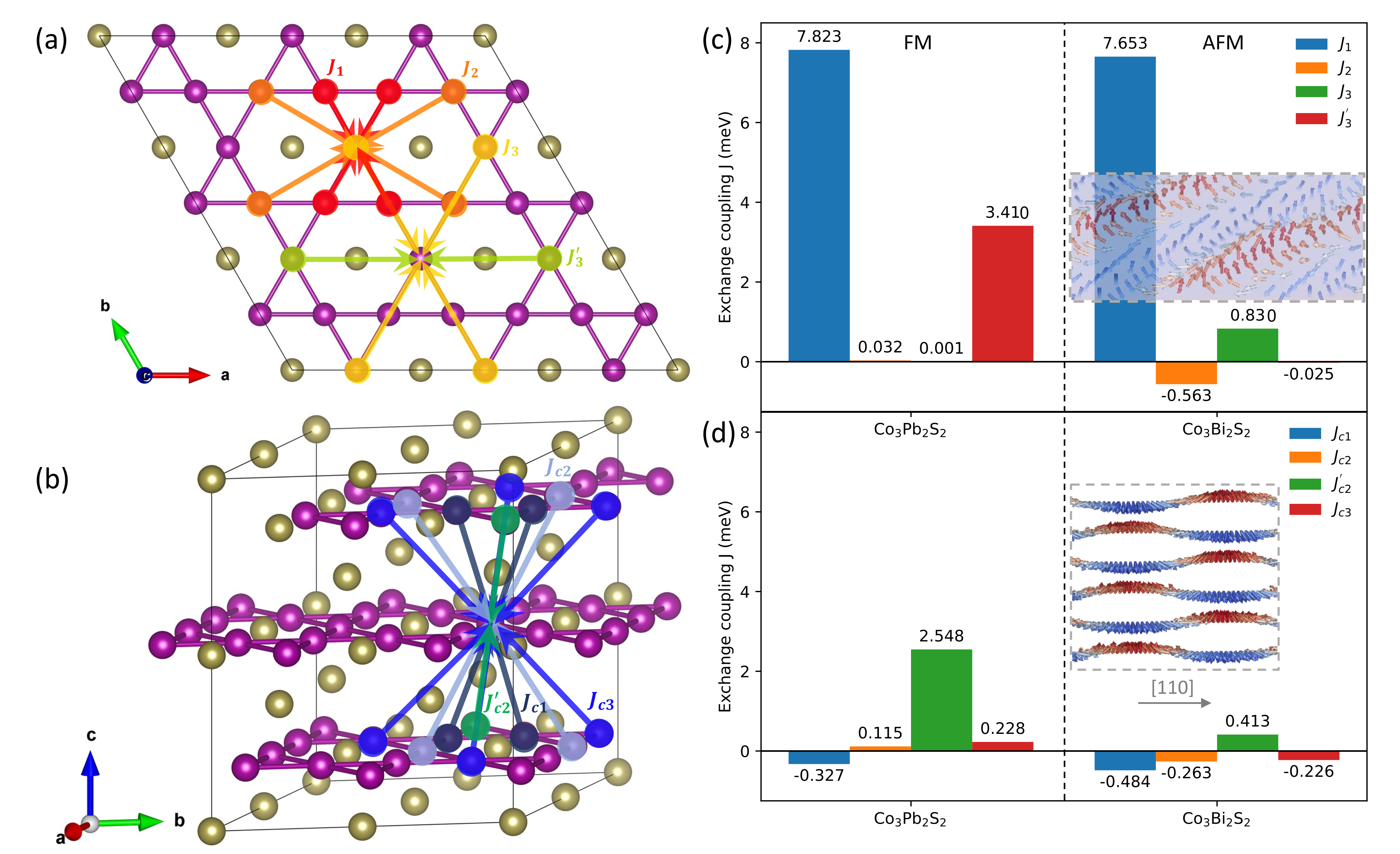}
		\caption{Illustration of $J_{ij}$ pairs corresponding to (a) in-plane and (b) out-of-plane exchange couplings. In a comparative manner between Co$_3$Pb$_2$S$_2$ and Co$_3$Bi$_2$S$_2$ which are predicted to show FM and AFM ground stats, respectively, the magnitudes of the dominant in-plane and out-of-plane exchange coupling parameters are demonstrated in (c) and (d), respectively. The insets of (c) and (d) show the spin-spiral like spin configurations viewed from in-plane and out-of-plane perspectives, respectively, in AFM Co$_3$Bi$_2$S$_2$.}
		\label{fig:jij_str}
	\end{center}
\end{figure*}

\subsection{Poisson's ratio} 
Poisson's ratio describes the plasticity behavior of the material, which measures the deformation in the direction perpendicular to the applied internal forces. When a stretching force is applied in the longitudinal direction, a material initiates to elongate. Simultaneously, the cross-section contracts in the transverse direction. This results in a positive value for Poisson's ratio, which is the ratio of transverse contraction to longitudinal elongation. The linear Poisson's ratio can be calculated directly from the bulk and shear modulus values, using the formula $\nu$ = (3$\it B$-2$\it G$)/(2(3$\it B$+$\it G$)). For HAP compounds with $\Delta E_H=0$, the positive Poisson's ratio varies between 0.11 in Co$_3$Al$_2$S$_2$ and 0.30 in Co$_3$Al$_2$S$_2$ (Table~\ref{tab:table1}).\par
However, a negative Poisson's ratio is obtained when the cross-section of materials elongates in response to the longitudinal stretching force. Such materials are called auxetic materials. Auxetics materials have numerous applications, such as sensor technology, air filters, fasteners, soundproof materials and high-performance armor~\cite{greaves2011poisson,lakes1987foam,chen1991holographic,alderson1999triumph,evans2000auxetic}. The negative Poisson's ratio is quite a rare phenomenon compared to the positive Poisson's ratio. Baughman et al. pointed out that the auxetic behavior is often observed in cubic elemental metals when stretched along the [110] direction.~\cite{baughman1998negative} Our Poisson's ratio analysis found six HAP compounds with negative Poisson's ratio (see Table~\ref{tab:table1} and SI Table S2). For instance, Co$_3$Ge$_2$Se$_2$ and Ni$_3$Hg$_2$Se$_2$ exhibit a large negative Poisson's ratio of 0.10 and 0.21, respectively. The mechanism responsible for the origin of the auxetic effect in HAP materials is saved for future studies.
\begin{figure*}[htp]
\centering
	\begin{center}
	    \includegraphics[width=1.0\textwidth]{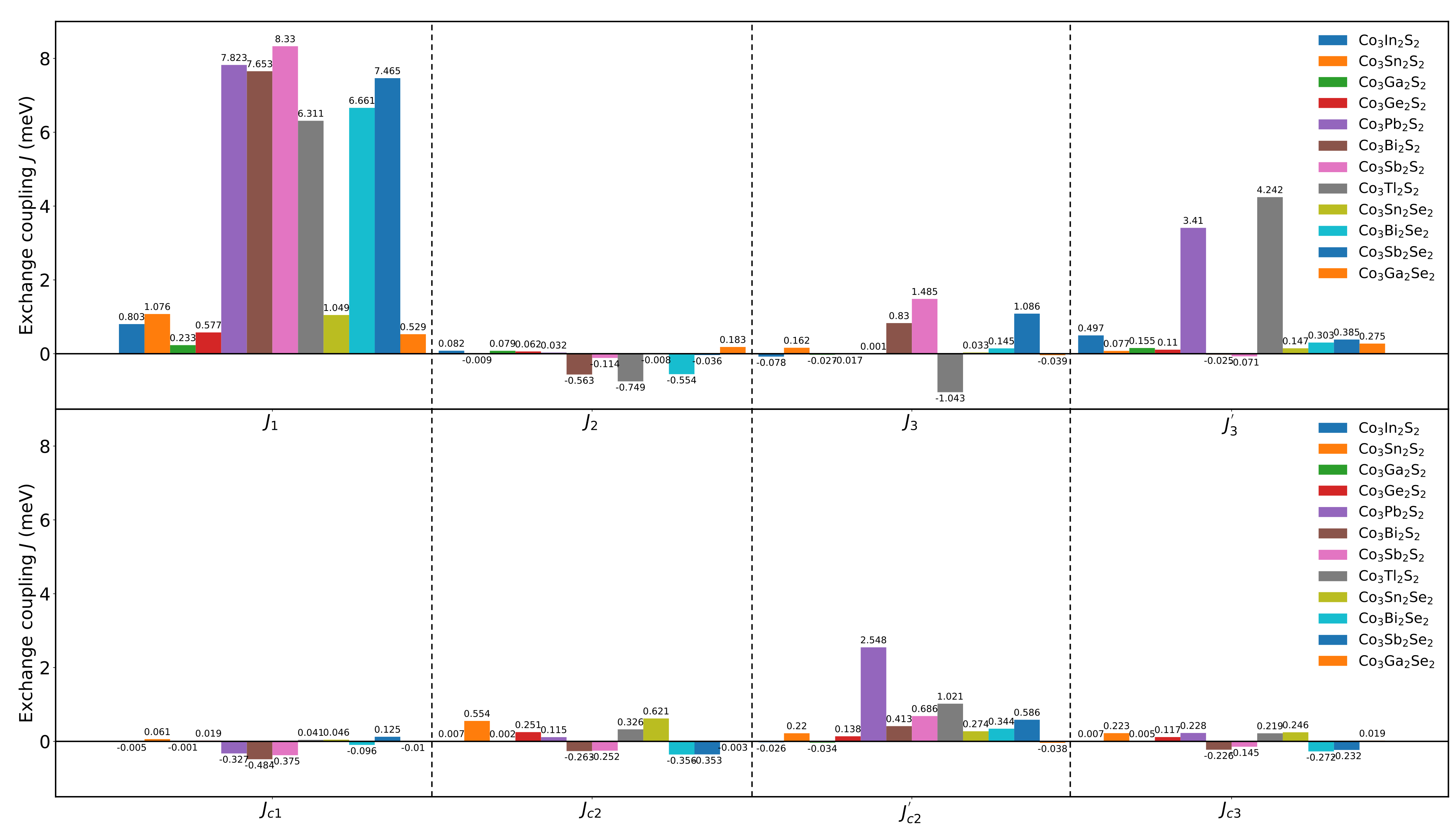}
		\caption{Upper panel: the in-plane exchange coupling parameters ($J_1$, $J_2$, $J_3$, an $J_{3}^{\prime}$) in stable Co-based HAPs; Lower panel: the out-of-plane exchange coupling parameters ($J_{c1}$, $J_{c2}$), $J_{c2}^{\prime}$ and $J_{c3}$) in stable Co-based HAPs}
		\label{fig:J_Co}
	\end{center}
\end{figure*}

\subsection{Magnetic ground state and transition temperature}
We list in Table~\ref{tab:table1} the MGSs predicted by DFT and ASD simulations, respectively. The obtained critical magnetic order transition temperatures via the MF theory are also provided. 
First, the calculated magnetic exchange parameters and the MF $T_C$ of Co$_3$Sn$_2$S$_2$ are compared to previous work, as summarized in Table~\ref{tab:mag}).~\cite{zhang2021unusual,liu2021spin}
For clarification of the magnetic exchange pairs, we illustrate in Figure~\ref{fig:jij_str}(a) and (b) the first three nearest neighbors corresponding to the in-plane and out-of-plane interactions, respectively. Note that the third-nearest neighbors for the in-plane exchange pairs can be classified into two groups, one of which crosses element X1 (marked by $J_3^{\prime}$, see Figure~\ref{fig:Structure} (a) for the notation of X1). Similarly, for the out-of-plane exchange pairs, the second-nearest neighbors also consists of two different groups, in which one group sees element X2 in between (marked by $J_{c2}^{\prime}$, see Figure~\ref{fig:Structure} (a) for the notation of X2).

As demonstrated in Table~\ref{tab:mag}, the so-obtained exchange parameters $J_1$, $J_2$, $J_3$, $J_3^{\prime}$, $J_{c1}$, $J_{c2}$, $J_{c2}^{\prime}$, $J_{c3}$ are of 1.08, -0.01, 0.16, 0.08, 0.06, 0.55, 0.22, 0.23 meV, respectively, and the corresponding MF $T_C$ is around 81 K for Co$_3$Sn$_2$S$_2$. In comparison with the results reported in Ref.~\cite{zhang2021unusual}, which adopted the SPR-KKR code for the GGA + U calculations,~\cite{ebert2011calculating} the magnitudes of $J_1$, $J_3^{\prime}$ and $J_{c3}$ calculated using the GGA + U approach are larger. 
In fact, both OpenMX and SPR-KKR calculate the magnetic exchange parameters based on the magnetic force theorem.~\cite{liechtenstein1987local} For a more direct comparison, we also carried out a GGA + U calculation using OpenMX with the same $U$ (1.1 eV) and $J$ (0.1 eV) applied on Co 3$d$ orbitals as in Ref.~\cite{zhang2021unusual}. It can be observed that with $U$ applied on Co site, the magnitude of $J_1$ is nearly doubled (2.12 meV versus 1.08 meV) and the value of $J_{c2}$ is increased. Consequently, the predicted MF $T_C$ also gets higher (127 K) as compared to the bare GGA simulation (81 K). Although the predicted $T_C$ using OpenMX together with GGA + U (127 K) is quite close to the $T_C$ given by the SPR-KKR code (130 K), the magnitudes of the exchange interaction parameters are not totally consistent, especially for $J_1$ and $J_3^{\prime}$. 
The OpenMX with GGA + U method gives in general better agreement when compared to the results of Ref.~\cite{liu2021spin} which calculated the exchange parameters by employing a combination of magnetic force theorem and Wannier functions approach.~\cite{korotin2015calculation} However, one should be reminded that it is always a question regarding what are the most appropriate $U$ and $J$ values. 
In addition, the fitted exchange interactions based on the experimental dispersions and intensities of spin-wave (SW) spectra, as well as the experimental $T_C$, are listed in Table~\ref{tab:mag} for reference.
Despite the deviations present in the specific values of exchange couplings, the comparison among different \textit{ab} \textit{initio} methods indicates that the bare OpenMX method predicts resonably well the critical magnetic order transition temperature of a HAP.

Furthermore, the ASD method was utilized to cross check the obtained MGSs using the DFT approach. Discrepancies on the obtained MGSs by DFT and ASD approaches can be identified in Table~\ref{tab:table1}. For instance, the ASD simulation gives AFM as the MGS of Co$_3$Bi$_2$S$_2$, in contrast to the FM state given by DFT. Such discrepancy is owing to the limits of DFT calculations on running through all possible AFM configurations. And the complexity of properly determining the MGS with a non-collinear AFM pattern is discussed in the following texts. 
We demonstrate in Fig~\ref{fig:jij_str} (c) and (d) the dominating in-plane and out-of-plane magnetic exchange coupling parameters, respectively, of two cases: Co$_3$Pb$_2$S$_2$ (predicted to possess FM ground state using both DFT and ASD) versus Co$_3$Bi$_2$S$_2$ (predicted to possess FM ground state using DFT whereas AFM ground state using ASD). For both Co$_3$Pb$_2$S$_2$ and Co$_3$Bi$_2$S$_2$, the in-plane $J_1$ is the dominant exchange interaction. In addition, $J_3^{\prime}$ and the out-of-plane $J_{c2}^{\prime}$ of Co$_3$Pb$_2$S$_2$ exhibit moderate magnitudes of around 3.410 and 2.548 meV, respectively, with only one exception that $J_{c2}$ is slightly negative ($\sim$-0.327 meV). As a consequence, Co$_3$Pb$_2$S$_2$ is stabilized at the FM state, which is verified by ASD simulation as well. In contrast to Co$_3$Pb$_2$S$_2$, Co$_3$Bi$_2$S$_2$ prefers AFM to FM according to the ASD simulation. We show in the inset of Figure~\ref{fig:jij_str}(c) and (d) the predicted in-plane and out-of-plane spin distributions of Co$_3$Bi$_2$S$_2$ at 10 K, respectively, from which one can clearly see an AFM spin-spiral state on the (001) surface and along the [110] direction. Moreover, it is noticed that the Co site in Co$_3$Bi$_2$S$_2$ aligns in an AFM manner along the \textbf{\textit{c}} direction. 
Based on the $J$ parameters, we speculate that such spin configuration is preferred mainly due to the negative $J_{2}$ and $J_{c1}$ couplings for Co$_3$Bi$_2$S$_2$. In analogous to Co$_3$Bi$_2$S$_2$, the disagreements in DFT and ASD predictions on the MGSs of Co$_3$Sb$_2$S$_2$, Fe$_3$In$_2$S$_2$, Mn$_3$In$_2$S$_2$, Mn$_3$Sb$_2$S$_2$, Mn$_3$Sn$_2$S$_2$, Co$_3$Bi$_2$Se$_2$, Co$_3$Ga$_2$Se$_2$, Mn$_3$Sb$_2$Se$_2$ and Mn$_3$Sn$_2$Se$_2$ can be interpreted as the relatively stronger competitions between the FM and AFM exchange interactions therein (see SI Table S3 for the listed exchange parameters). 
At last, as mentioned in Sec.~\ref{sec:method:MC}, for the HAPs that are predicted to have AFM ground states by both DFT and ASD approaches (a majority of them are Mn-based as shown in Table~\ref{tab:table1}), the AFM state given by ASD simulation is, as a matter of fact, different from the one shown in Figure~\ref{fig:Structure} (b). A more thorough study on the non-collinear AFM state for particularly Mn-based HAPs will be the main subject of our future work. 

Therefore, the complexities in predicting the true MGSs of HAPs are hinted in the above comparative discussions of Co$_3$Pb$_2$S$_2$ and Co$_3$Bi$_2$S$_2$. The diversity of the magnetic exchange couplings motivates us to gain a more complete picture of their behaviors in a group of, \textit{e.g.}, Co-based or Mn-based HAPs. 
We summarize in Figure~\ref{fig:J_Co} the distributions of exchange coupling parameters for all the magnetic Co-based HAPs as listed in Table~\ref{tab:table1}. The value of $J_1$ is positive for all the Co-based HAPs listed here. The first-nearest neighbor out-of-plane exchange interaction $J_{c1}$ is rather weak and a negative sign is usually seen. We also notice that for Co$_3$Pb$_2$S$_2$ and Co$_3$Tl$_2$S$_2$, the exchange parameters $J_{3}^{\prime}$ and $J_{c2}^{\prime}$, whose exchange pairs cross Pb/Tl atom, exhibit fairly strong FM couplings. As a result, Co$_3$Pb$_2$S$_2$ and Co$_3$Tl$_2$S$_2$ are predicted to possess a FM state with $T_C$ of approximately 329 and 266 K, respectively (see Table~\ref{tab:table1}). Co$_3$Bi$_2$S$_2$ and Co$_3$Bi$_2$Se$_2$ belong to another category, in which $J_{2}$ and $J_{c1}$ indicate a stronger preference to AFM order. Hence the ASD simulations predict AFM being their MGSs. 
Note that all the exchange interactions of Co$_3$In$_2$S$_2$ are in general weak, which could explain why a paramagntic state was proposed~\cite{mcguire2021antiferromagnetic}.
In general, the magnetic frustrations are highly expected in HAPs containing the kagome lattice. The opposite sign of $J_1$ and $J_2$ can be observed in several materials, as listed in SI Table S3, besides the potential in-plane AFM order implied by the negative $J_{c1}$. 

\section{Conclusions}
In summary, a systematic high-throughput screening was performed on 930 HAP compounds to search for novel stable magnetic systems. After validating the stability of 6 experimentally reported compounds based on the thermodynamical and mechanical stability criteria, we found 45 stable novel compounds. The number of stable compounds rises to 102 if the convex hull tolerance ($\Delta E_H$) is increased to 100 meV/atom. Further COHP analysis suggests that M-Z bonds contribute most to the thermodynamic stability.
Subsequently, we invesgated various mechanical properties such as bulk modulus ($\it B$), shear modulus ($\it G$), Young's modulus ($\it E$), and Poisson's ratio ($\it v$) for the thermodynamically stable compounds. We found six compounds with negative Poisson's ratio, which is an uncommon occurrence and can be useful for applications like sensors. Moreover, the systematic calculations of the magnetic exchange coupling parameters indicate that it is difficult to properly determine the MGS by employing solely the DFT method. Instead, by adopting the ASD simulations, the predicted MGSs by DFT can be further verified or revised. 
In the end, we show that 23 compounds would host a non-collinear AFM state and 14 of them show a $T_N$ higher than room temperature. A number of 11 HAPs were found to possess FM ground state and 5 of them exhibit a $T_C$ higher than room temperature. The investigations on the MGS and $T_C/T_N$ provide a valuable guidance for designing novel magnetic materials for spintronics applications.
\section*{Acknowledgements}
The authors are grateful and acknowledge TU Darmstadt Lichtenberg's high-performance computer (HPC) support for the computational resources where the calculations were conducted for this project. This project was supported by the Deutsche Forschungsgemeinschaft (DFG, German Research Foundation)-Project-ID 405553726-TRR 270.

\nocite{*}
\clearpage
\bibliography{Acta}
\end{document}